# Spin and orbital moments of nanoscale Fe$_3$O$_4$ epitaxial thin film on MgO/GaAs(100)

W. Q. Liu[1,2], Y. B. Xu[1,2,*], P. K. J. Wong[2], N. J. Maltby[2], S. P. Li[2], X. F. Wang[1], J. Du[1,3], B. You[3], J. Wu[4], P. Bencok[5], R. Zhang[1*]

[1] York-Nanjing International Center of Spintronics, School of Electronics Science *and Engineering*, Nanjing University, Nanjing 210093, China

[2] Spintronics and Nanodevice Laboratory, Department of Electronics, University of York, York YO10 5DD, UK

[3] National Laboratory of Solid State Microstructure, Department of Physics, Nanjing University, Nanjing 210093, China

[4] Department of Physics, University of York, York YO10 5DD, UK

[5] Diamond Light Source, Didcot OX11 0DE, United Kingdom

**ABSTRACT:** Nanoscale Fe$_3$O$_4$ epitaxial thin film has been synthesized on MgO/GaAs(100) spintronic heterostructure, and studied with X-ray magnetic circular dichroism (XMCD). We have observed a total magnetic moment ($m_{l+s}$) of $(3.32\pm0.1)\mu_B$/f.u., retaining 83% of the bulk value. Unquenched orbital moment $(m_l)$ of $(0.47\mu_B\pm0.05)\mu_B$/f.u. has been confirmed by carefully applying the sum rule. The results offer direct experimental evidence of the bulk-like total magnetic moment and a large orbital moment in the nanoscale fully epitaxial Fe$_3$O$_4$/MgO/GaAs(100) heterostructure, which is significant for spintronics applications.

In the contemporary spintronics research, magnetite, especially in nano-form, has attracted enormous interest due to its great potential for information technology. To synthesize highly spin polarized materials as spin sources and to combine them with semiconductors (SCs), which can be easily integrated with current magnetic technologies are prerequisites for proposed spintronics devices such as the spin field effect transistor (SFET).[1] As a model half-metallic material, $Fe_3O_4$ has shown 80% spin polarization near the Fermi level in experiment[2] and theoretically, up to 100% can be expected[3]. More desirably, the high Curie temperature of $Fe_3O_4$ makes it a promising candidate for room temperature use. Fascinating properties of spin transport have also been presented in $Fe_3O_4$, i.e. spin Seebeck effect[4], spin filter effect[5], gate voltage-induced phase transition[6], and spin valve effect of $Fe_3O_4$/MgO/$Fe_3O_4$ junctions[7]. Yet at the meantime, many fundamental properties of magnetite such as the half-metallicity, spin and orbital ordering, Verwey transition mechanism and the coupling mechanism between different sites have long been open issues, and with the thickness down to nanometer scale, these issues become even more sophisticated.

Magnetite is believed to have a cubic inverse spinel structure, where $Fe^{3+}$ ions occupy tetrahedral sites (usually called A sites), whereas octahedral sites (B sites) are occupied by both $Fe^{3+}$ and $Fe^{2+}$ ions. The spin of $Fe^{3+}$ ions at octahedral and tetrahedral sites are aligned antiparallel to each other leading to a net spin magnetic moment ($m_s$) of 4 $\mu_B$/f.u., corresponding to a fully occupied local majority band (opposite for A and B sites). The presence of integer magnetic moment of magnetite is expected in experiment as indication for a B-site minority electron conduction mechanism, and its accompanied full spin polarization at the Fermi level. However, controversial results of it have been reported based on techniques including superconducting quantum interference device (SQUID) magnetometer[8,34], XMCD[9,21,25,35, 34], and magnetic Compton scattering (MCS)[28,29] and calculations including local density approximation (LDA)[25], LDA+U[25], local spin density approximation (LSDA)+U[3], and moment analysis[9] etc. The reported $m_l$ of magnetite varies from (0.67±0.09) $\mu_B$/f.u. by Huang *et al*.[25], (0.51±0.05) $\mu_B$/f.u. by Li *et al*.[28], to (0.06±0.14) $\mu_B$/f.u. by Duffy *et al*.[29] and all the way down to –0.001$\mu_B$/f.u. by Goering *et al*.[21]. Theoretical analyses were given with even sharper contrast, varying from 0.43 $\mu_B$/f.u. by Huang *et al.*[25] to 0.02 $\mu_B$/f.u. by Antonov *et al.*[3]. Similar controversial observations exist for the $m_s$ of magnetite. Goering *et al.*[35] reported $m_s$ down to (1.7±0.02) $\mu_B$/f.u in single crystal $Fe_3O_4$. Among the very few work on $Fe_3O_4$ thin films, Orna *et al.*[34] observed greatly reduced $m_s$ = 1.83 $\mu_B$/f.u. of $Fe_3O_4$ on MgO, as well as Babu *et al.*[27] observed $m_s$ = (1.20±0.05) $\mu_B$/f.u. of $Fe_3O_4$ on $BaTiO_3$. On the other hand, enhanced $m_s$ = 7.7 $\mu_B$/f.u. was reported by Arora *et al.*[8]. The well-known Verwey transition of magnetite is believed to be accompanied by a transition to a low symmetry structure, across which, the $m_{l+s}$ and especially $m_l$ are expected to change significantly. The experimental work[8,29,34], however, has so far found no difference of them across the transition, which questions the picture of a fully A site $Fe^{3+}$ and a mixed-valence B site configuration of magnetite.

Direct epitaxial growth of $Fe_3O_4$ on GaAs(001) by *in situ* postgrowth annealing was firstly achieved by Lu *et al.*[10]. Beyond this classic magnetic/semiconductor (FM/SC) heterostructures, $Fe_3O_4$/MgO/GaAs system could be a rather more timely and important system to look into, as crystalline MgO-based magnetic tunnel junctions have achieved great success in assisting efficient spin injections for various applications[11,12,13]. An insulating layer of MgO can be used as a tunnelling barrier, which not only relieves the conductivity mismatch problem but also works as a spin filter. Moreover, MgO forms an excellent diffusion barrier with thermal stability up to 800ºC, effectively preventing the intermixing at a given ferromagnet-semiconductor interface. Efforts have been made to explore the magnetic behaviour of nanoscale epitaxial thin films on bulk MgO(100)[8]. Yet to fundamentally understand its electronic nature, e.g. the character of 3d electrons in $Fe_3O_4$, one needs to extract the spin and orbital magnetic moments of $Fe_3O_4$ respectively and such challenging task forms the core of our study. In this work, we present systematic XMCD measurements of $Fe_3O_4$ epitaxial thin film on MgO/GaAs(001), aiming to contribute to the open question of the magnetic moments of the $Fe_3O_4$ ultrathin film.

The magnetite thin film we have studied here was grown by post-annealing of an epitaxial Fe(001) in an oxygen partial pressure. Firstly a sharp GaAs(001) surface was obtained after annealed in an ultrahigh vacuum chamber for 40 minutes at 830 K as identified from the reflection high-energy electron diffraction (RHEED) pattern shown in Fig. 1(a). The 1nm MgO buffer layer was then grown by e-beam evaporation at a rate of 2 Å/min, as monitored by a quartz microbalance calibrated by ex-situ AFM, while the substrates were kept at 673 K. The chamber pressure during deposition was below 4 x $10^{-8}$ mbar suggesting a limited decomposition of the MgO crystals. Lastly, a 4 nm thick epitaxial Fe film was grown on the MgO/GaAs at room temperature, followed by post-growth annealing at 500 K in the oxygen partial pressure at 5 x $10^{-5}$ mbar for 10 minutes. Overall the comparatively large oxygen ions form an face-centered-cubic lattice and the Fe atoms are located in interstitial sites. The epitaxial relationship can be identified from RHEED as $Fe_3O_4$(100)[001]//MgO(100)[001]//GaAs(100)[001] as shown in Fig. 1(a)-(d), whose evolution during the growth has been discussed by Wong *et al*. elsewhere[14].

The general magnetic properties and stoichiometry of the sample were characterized *ex situ* by means of SQUID. The magnetization versus field (M-H) loops in Fig. 1(e) were obtained by applying the magnetic field out of plane. A small diamagnetic contribution from the sample holder has been carefully subtracted from the measured data. It can be seen that the sample exhibits a magnetization ~ 400 emu/$cm^3$ at 4 T, slightly lower than the previous reported value of 480 emu/$cm^3$ for bulk magnetite. It is generally accepted that the presence of Verwey transition is very sensitive to the stoichiometry and homogeneity of magnetite thin films[15]. The temperature ($T_V$) at which such a transition occurs has been commonly observed to decrease from the bulk value down to 85K and could even disappear with the decreasing $Fe_3O_4$ thicknesse[8,16,17,18,34]. Fig. 1(f) presents the temperature dependence of the magnetization of our sample, which was done by cooling the sample from 300 to 4.3 K in zero magnetic field, followed by an application of a static magnetic field of 100 Oe and recording the magnetization values during the warming cycle to 300 K. The magnetization drops with the decreasing temperature and a Verwey transition at ~ 100 K can be distinguished from dM/dT (see the insert of Fig. 1(f)). This is highly consistent with the reported $T_V$ of the magnetite films on or near stoichiometry[8,19], indicating the sample is not suffering to a great extent from cation or anion vacancies.

X-ray absorption spectroscopy (XAS) and XMCD experiments at the Fe $L_{2,3}$ absorption edges were performed at beamline I10 of the UK National Synchrotron Radiation Laboratory. The XAS experiments were carried out at 300 K under an applied field ranging from 1T to 10 T with total electron yield (TEY) detection. Circularly polarized X-rays with 100% degree of polarization were used in normal incidence with respect to the sample plane and in parallel with the applied magnetic field, so as to minimize the nonmagnetic asymmetries. The XMCD was obtained by taking the difference of the XAS spectra, e.g. $\sigma^+$-$\sigma^-$, achieved by flipping the X-ray helicity at a fixed magnetic field. As shown in Fig. 2, the observed XMCD is similar to those of the reported $Fe_3O_4$ spectra caused by the antiparallel spin orientations of A and B sites. The B sites $Fe^{3+}$ and $Fe^{2+}$ spin-up states exhibit negative peaks at Fe $L_3$ edge and positive peaks at the Fe $L_2$ edge, while the A sites $Fe^{3+}$ spin-down states behave the oppositely at the Fe $L_3$ and $L_2$ edges, respectively. Similar features of the XMCD line shape were also observed at enhanced magnetic fields up to 10T, as presented in Fig. 3.

The $m_s$ and $m_l$ were calculated by applying sum rules on the integrated XMCD and total XAS spectra of Fe $L_{2,3}$ edges based on equation (1)[20], where $n_h$ - the effective number of 3d-band holes had been taken from literature[25]. In order to rule out non-magnetic parts of the XAS spectra an arctangent based step function is used to fit the threshold[20]. As can be seen from Fig. 2, unlike an infinite 'long tail' reported by Goering *et al*.[21], the integrated spectrum of both XMCD and total XAS of our data quickly saturate at ~ 728 eV. Therefore an integration range to 735 eV is sufficient, giving $m_s$ = (2.84±0.1) $\mu_B$/f.u., $m_l$ = (0.47±0.05) $\mu_B$/f.u. and the $m_l/m_s$ ratio as large as 0.18. It should be noted that all sum rule-related values given here are average information over the whole formula unit (f.u.) of all three Fe ions. Besides, no corrections have been applied due to an incompleted circular polarization of the X-ray, given two APPLE II undulators are used on the beamline which generate linear/elliptical polarization at any arbitrary angle from 0° to 180°[22]. Possible artifact of the experimental set up and data analysis of XMCD of magnetite were discussed in detail by Goering *et al*.[21] In general, the nonmagnetic part of the raw data is smaller than 1/1000 of the total absorption. The saturation effect in our case is estimated to be about 3% in the normal incidence configuration. Besides, the magnetic dipole term <Tz> plays a rather insignificant role because of the predominantly cubic symmetry of magnetite, even under a scenario of additional surface symmetry breaking. The good agreement of the total magnetic moment obtained from SQUID measurement (400 emu/cm$^3$) with the $m_{l+s}$ = (3.32±0.1) $\mu_B$/f.u. calculated from XMCD is an additional proof of the proper application of the sum rule in our study.

$$m_l = -\frac{4}{3} n_d \frac{\int_{L_{2,3}} (\sigma^+ - \sigma^-) \, dE}{\int_{L_{2,3}} (\sigma^+ + \sigma^-) \, dE}$$

$$m_s + <Tz> = -n_d \frac{6\int_{L_3} (\sigma^+ - \sigma^-) \, dE - 4\int_{L_{2,3}} (\sigma^+ - \sigma^-) \, dE}{\int_{L_{2,3}} (\sigma^+ + \sigma^-) \, dE}$$

(1)

Unquenched $m_l$, or strong spin-orbit coupling (<LS>), is a desired property in terms of the controllability by electric field in spintronics operation[23], however, which have been reported with controversy in magnetite. Early theoretical work based on the picture of bulk Fe$_3$O$_4$ possessing $m_s$ = 4 $\mu_B$/f.u. and nearly vanishing $m_l$. McQueeney *et al.*[24] obtained a high <LS> of magnetite of the order of 10 meV, pointing to a large $m_l$ to expect. The XMCD performed by Huang *et al.*[25] indicated a large unquenched $m_l$, typically 0.67 $\mu_B$/f.u. along with a spin moment 3.68 $\mu_B$/f.u. at the temperature both above and below Verwey transition. These results are consistent with the $m_s$ and $m_l$ which calculated by them using the LDA + U scheme[25]. The large $m_l$ has been attributed to a strong on-site Coulomb interaction and corresponding 3d correlation effects. Similarly sizable $m_l$ was also observed by Kang *et al.*[26] in Mn substitution at the A site, which changes the valence of the B-site Fe and by Babu *et al.*[27] in ultrathin Fe$_3$O$_4$ on BaTiO$_3$(001). By sharply contrast, XMCD performed by Goering *et al.*[21] suggests that there is in fact a vanishingly small $m_l$ on the Fe sites. To avoid the systematic errors arises from the XMCD data analysis, MCS were performed, which still end in controversial results. Non-integral $m_s$ = 3.54 $\mu_B$/f.u. and correspondingly $m_l$ = 0.51 $\mu_B$/f.u. were observed by Li *et al.*[28] while Duffy *et al.*[29] reported again nearly vanishing $m_l$. Goering *et al.*[9] has recently tried to explain the large variety of published results by the independent analysis of the Fe $L_{2,3}$ edge XAS, moment analysis fit of the Fe $L_{2,3}$ edge XMCD, and by the comparison with O K edge XMCD. In consistent with Goering's, our data also exhibit an intensity ratio $r_{23}=0.25$, strongly reduced from a pure statistical case where $r_{23}=0.5$. This is a clear proof for the presence of large $m_l$ and respective <LS> expectation values in our ultrathin $Fe_3O_4$ film. According to Goering's argument, these orbital moments are located at the A and B sites of magnetite and aligned antiparallel with each other, similar to the spin moments, though quantitatively, this scenario questions the picture of a fully A site $3^+$ and a mixed-valent B site configuration. Table 1 summaries some of the experimental and theoretical efforts toward this issue. Regardless the controversial reports on various form of magnetite, our results (the first line of the table) confirms the existence of a large unquenched $m_l$ with $m_l/m_s = 0.17$ in the ultrathin $Fe_3O_4$ film grown on the MgO/GaAs(100). While the unquenched $m_l$ might be a intrinsic property of the bulk $Fe_3O_4$, which is still a hotly-debated topic, our result could also originates from modification of crystal lattice symmetry as $m_l$ in the low dimensional magnetic systems can be strongly enhanced by the reduced symmetry of the crystal field as found in the Fe/GaAs(100) system[30].

　　　It is worth noting that not only the nanoscale full epitaxial $Fe_3O_4$/MgO/GaAs(100) heterostructure exhibits considerably large $m_l$, its $m_{l+s}$ retains about 87% of the bulk value. The deviation of magnetic moments of epitaxial thin films from the bulk value is usually attributed to three forms of missing compensation or symmetry breaking. The first one is the formation of antiphase boundaries (APBs) raised from the epitaxy growth process due to the fact that $Fe_3O_4$ has twice the unit-cell size of MgO[31,32]. In magnetite thin films, the magnetic interactions are altered at the APBs, across which the intrasublattice exchange interactions dominate, reversing the spin coupling. Therefore, the structural boundary separates oppositely magnetized regions and the resultant coupling between two domains turns out to be either frustrated or antiferromagnetic. To exclude the presence of APBs, we repeated the experiment at enhanced magnetic fields up to 10T, since such antiferromagnetic exchange interactions usually lead to large saturation fields[33]. As plotted in Fig. 4, a rather consistent value of $m_s$ and $m_l$ have been extracted at saturation from 4-10T within the error bar, which rules out any significant effects caused by APBs, while the value of $m_{l+s}$ obtained at 1T is slightly smaller because of unsaturation as expected. The second mechanism of non-compensation occurs due to the less cubic symmetry of magnetite at the surface and the interface of $Fe_3O_4$/MgO. Among the very few work performed on $Fe_3O_4$ thin films, Orna et al.[34] reported significantly shrinking $m_s$ = 1.83 $\mu_B$/f.u. in $Fe_3O_4$(8nm)/MgO, as well as the observation by Babu et al.[27] of $m_s$ = (1.20±0.05) $\mu_B$/f.u. in $Fe_3O_4$(2.5nm)/$BaTiO_3$. Even in the bulk, strongly reduced $m_s$ of $Fe_3O_4$ was also observed by Goering et al.[35] down to (1.73±0.02) $\mu_B$/f.u.. By contrast again, large $m_s$ of 7.7 $\mu_B$/f.u in $Fe_3O_4$(5nm)/MgO was reported by Arora et al.[8], who attributed the enhancement to the uncompensated spin between A and B sublattices at the surface and across the APBs. However, this enhancement may also come from the magnetic impurity as suggested by Orna et al.[34]. The inter-diffusion of $Mg^{2+}$ ions, which tends to substitute onto B-sites, is the third possibility. Although our sample were grown at a moderate growth temperature (500K), one may still predict an appreciable inter-diffusion given the *ex situ* measurement in the study were not carried out immediately after the growth. However, if any, such substitution would only happen at the first 1-2 atom layers at the $Fe_3O_4$/MgO interface. Therefore figures in this paper are more representative for a $Fe_3O_4$ thin film on MgO/GaAs with consistent stoichiometry. Nevertheless, the bulk-like magnetic moment as found in our work indicates that the $Fe_3O_4$ ultrathin film synthesized by post-growth annealing under 500K can effectively prevent the formation of APBs and the interdifussuon of $Mg^{2+}$ ions into the magnetic layer.

　　　　To summarize, we have performed XMCD of a $Fe_3O_4$ epitaxial thin film on MgO/GaAs(100) synthesized by posting-growth annealing. High quality of XAS and XMCD spectra were obtained and carefully analyzed with the sum rule. A significant unquenched $m_l$ was found, which may come from the breaking of the symmetry at the interfaces in this low dimension system. The observed sizable $m_l$ has strong implications for realizing spintronics operations as a high <LS> coupling is essential for the ultrafast switching of spin polarization by electric field and circularly polarized light. Moreover, unlike the reported significantly reduced values of the magnetic moment of $Fe_3O_4$ ultra-thin films, our $Fe_3O_4$/MgO/GaAs(100) heterostructure retains a large $m_{l+s}$ of 83% of the bulk value. Based on the clear Verwey transition and the consistent magnetization values obtained at high magnetic field, we believe such large magnetic moments stay independent from nonstoichiometric defects (cation or anion vacancies) and the APBs. Our results offer direct experimental evidence addressing the open issue of the spin and orbital moments of magnetite, particularly, in its epitaxial ultrathin film form, which is significant for achieving high efficient spin injection and electrical spin manipulation in the full epitaxial spintronic heterostructure.

**ACKNOWLEDGEMENTS**

　　This work is supported by the State Key Programme for Basic Research of China (Grants No. 2014CB921101), NSFC (Grants No. 61274102) and PAPD project, UK EPSRC and STFC. We acknowledge Jill S. Weaver for sharing the computation code of the sum rule calculation.

**REFERENCES**

[1] S. Datta and B. Das, Appl. Phys. Lett. **56**, 665 (1990).

[2] Y. S. Dedkov, U. Rudiger, and G. Guntherodt, Phys. Rev. B **65**, 064417 (2002).

[3] V. Antonov, B. Harmon, and a. Yaresko, Phys. Rev. B **67**, 024417 (2003).

[4] R. Ramos, T. Kikkawa, K. Uchida, H. Adachi, I. Lucas, M.H. Aguirre, P. Algarabel, L. Morellón, S. Maekawa, E. Saitoh, and M.R. Ibarra, Appl. Phys. Lett. **102**, 072413 (2013).

[5] Z.-M. Liao, Y.-D. Li, J. Xu, J.-M. Zhang, K. Xia, and D.-P. Yu, Nano Lett. **6**, 1087 (2006).

[6] J. Gooth, R. Zierold, J.G. Gluschke, T. Boehnert, S. Edinger, S. Barth, and K. Nielsch, Appl. Phys. Lett. **102**, 073112 (2013).

[7] H.-C. Wu, O.N. Mryasov, M. Abid, K. Radican, and I. V Shvets, Sci. Rep. **3**, 1830 (2013).

[8] S. Arora, H.-C. Wu, R. Choudhary, I. Shvets, O. Mryasov, H. Yao, and W. Ching, Phys. Rev. B **77**, 134443 (2008).

[9] E. Goering, Phys. Status Solidi **248**, 2345 (2011).

[10] Y.X. Lu, J.S. Claydon, and Y.B. Xu, Phys. Rev. B **233304**, 1 (2004).

[11] H. Idzuchi, S. Karube, Y. Fukuma, T. Aoki, and Y. Otani, Appl. Phys. Lett. **103**, 162403 (2013).

[12] D. Kumar, P.C. Joshi, Z. Hossain, and R.C. Budhani, Appl. Phys. Lett. **102**, 112409 (2013).

[13] Y. Pu, J. Beardsley, P.M. Odenthal, a. G. Swartz, R.K. Kawakami, P.C. Hammel, E. Johnston-Halperin, J. Sinova, and J.P. Pelz, Appl. Phys. Lett. **103**, 012402 (2013).

[14] P. Kwan. J. Wong, W. Zhang, Y. Xu, S. Hassan, and S.M. Thompson, IEEE Trans. Magn. **44**, 2640 (2008).

[15] J. P. Shepherd, J.W. Koenitzer, R. Aragon, J. Spalek, and J. M. Honig, Phys. Rev. B **43**, 8461 (1991).

[16] X.W. Li, A. Gupta, G. Xiao, and G. Q. Gong, J. Appl. Phys. **83**, 7049 (1998).

[17] W. Eerenstein, T. T. M. Palstra, S. S. Saxena, and T. Hibma, Phys. Rev. Lett. **88**, 247204 (2002)

[18] G.Q. Gong, A. Gupta, G. Xiao, W. Qian, and V. P. Dravid, Phys. Rev. B **56**, 5096 (1997).

[19] J. P. Shepherd, J.W. Koenitzer, R. Aragon, J. Spalek, and J. M. Honig, Phys. Rev. B **43**, 8461 (1991).

[20] C.T. Chen, Y.U. Idzerda, H.-J. Lin, N. V Smith, G. Meigs, E. Chaban, G.H. Ho, E. Pellegrin, and F. Sette, Phys. Rev. Lett. **75**, 152 (1995).

[21] E. Goering, S. Gold, M. Lafkioti, and G. Schütz, Europhys. Lett. **73**, 97 (2006).

[22] H. Wang, P. Bencok, P. Steadman, E. Longhi, J. Zhu, and Z. Wang, J. Synchrotron Radiat. **19**, 944 (2012).

[23] W.G. Wang, M. Li, S. Hageman, and C.L. Chien, Nat. Mater. **11**, 64 (2012).

[24] R.J. McQueeney, M. Yethiraj, W. Montfrooij, J.S. Gardner, P. Metcalf, and J.M. Honig, Phys. B Condens. Matter **385**, 75 (2006).

[25] D.J. Huang, C.F. Chang, H.-T. Jeng, G.Y. Guo, H.-J. Lin, W.B. Wu, H.C. Ku, a. Fujimori, Y. Takahashi, and C.T. Chen, Phys. Rev. Lett. **93**, 077204 (2004).

[26] J. S. Kang, G. Kim, H. J. Lee, D. H. Kim, H. S. Kim, J. H. Shim, S. Lee, H. Lee, J. Y. Kim, B. H. Kim, and B. I. Min, Phys. Rev. B **77**, 035121 (2008).

[27] V. Hari Babu, R.K. Govind, K.-M. Schindler, M. Welke, and R. Denecke, J. Appl. Phys. **114**, 113901 (2013).

[28] Y. Li, P. a. Montano, B. Barbiellini, P.E. Mijnarends, S. Kaprzyk, and a. Bansil, J. Phys. Chem. Solids **68**, 1556 (2007).

[29] J.A. Duffy, J.W. Taylor, S.B. Dugdale, C. Shenton-Taylor, M.W. Butchers, S.R. Giblin, M.J. Cooper, Y. Sakurai, and M. Itou, Phys. Rev. B **81**, 134424 (2010).

[30] J. S. Claydon, Y.B. Xu, M. Tselepi, J.A.C. Bland, G. van der Laan, Phys. Rev. Lett., 93 (3) JUL 16 2004.

[31] D.T. Margulies, F.T. Parker, M.L. Rudee, F.E. Spada, J.N. Chap- man, P.R. Aitchison, and A.E. Berkowitz, Phys. Rev. Lett. **79**, 5162 (1997).

[32] F.C. Voogt, T.T.M. Palstra, L. Niesen, O.C. Rogojanu, M.A. James, and T. Hibma, Phys. Rev. B **57**, R8107 (1998).

[33] D.T. Margulies, F.T. Parker, M.L. Rudee, F.E. Spada, J.N. Chapman, P.R. Aitchison, and A.E. Berkowitz, **953**, 2 (1997).

[34] J. Orna, P. a. Algarabel, L. Morellón, J. a. Pardo, J.M. de Teresa, R. López Antón, F. Bartolomé, L.M. García, J. Bartolomé, J.C. Cezar, and a. Wildes, Phys. Rev. B **81**, 144420 (2010).

[35] E. Goering, S. Gold, M. Lafkioti, and G. Schutz, J. Magn. Magn. Mater. **310**, e249 (2007)

**TABLES**

| Sample | Method | $m_l$(μ$_B$/f.u.) | $m_s$(μ$_B$/f.u.) | $m_{l+s}$(μ$_B$/f.u.) | $m_l/m_s$ |
|---|---|---|---|---|---|
| 8nm Fe$_3$O$_4$/MgO/GaAs(100) | XMCD | 0.47±0.05 | 2.84±0.1 | 3.32±0.1 | 0.17 |
| 5nm Fe$_3$O$_4$/MgO(001)[8] | SQUID | | | 7.7 | |
| 8nm Fe$_3$O$_4$/MgO(001)[34] | XMCD | | | 1.83 | <0.05 |
| 2.5nm Fe$_3$O$_4$/BaTiO$_3$(001)[27] | XMCD | 0.44±0.05 | 1.20±0.05 | 1.64 | 0.37 |
| single crystal Fe$_3$O$_4$ [25] | XMCD | 0.65±0.07 | 3.68±0.09 | 4.33±0.09 | 0.18 |
| single crystal Fe$_3$O$_4$ [28] | MCS | 0.51±0.05 | 3.54±0.05 | 4.05±0.05 | 0.14 |
| single crystal Fe$_3$O$_4$ [21] | XMCD | −0.001 | 3.90±0.09 | 4.2±0.09 | -0.00026 |
| single crystal Fe$_3$O$_4$ [35] | XMCD | <0.03±0.02 | 1.7±0.02 | <1.73±0.02 | <0.0018 |
| single crystal [29] | MCS | 0.06±0.14 | 4.08±0.03 | 4.14±0.14 | 0.03 |
| Theory[25] | LDA | 0.06 | 4.0 | 4.06 | 0.015 |
| Theory[25] | LDA+U | 0.43 | 4.0 | 4.43 | 0.108 |
| Theory[3] | LSDA+U | 0.02 | 3.7 | 3.72 | 0.005 |

**Table 1.** Spin and orbital moment of the magnetite of our sample and those from the literatures.

**Figures**

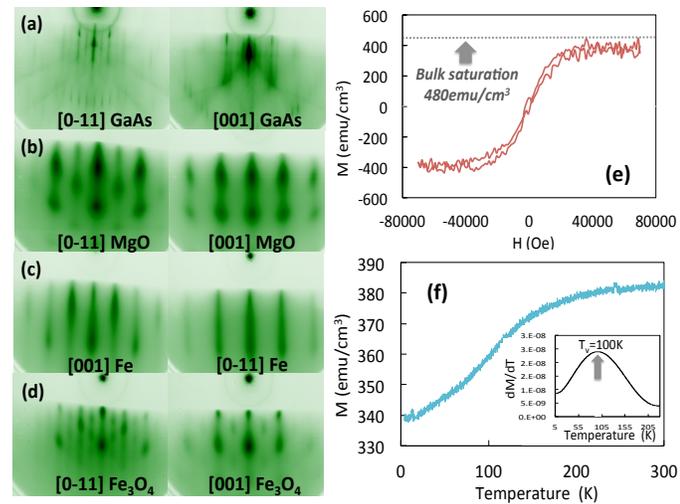

**Figure 1.** (a)-(d) RHEED patterns of (a) GaAs(100), (b) MgO/GaAs, (c) Fe/MgO/GaAs, and (d) $Fe_3O_4$/MgO/GaAs along [0-11] and [001] directions respectively, (e) The out of plane magnetization hysteresis loop of the sample. Dash line marks the value of bulk saturation from literatures, (f) Temperature dependence of the sample magnetization, insert: dM/dT v.s. temperature.

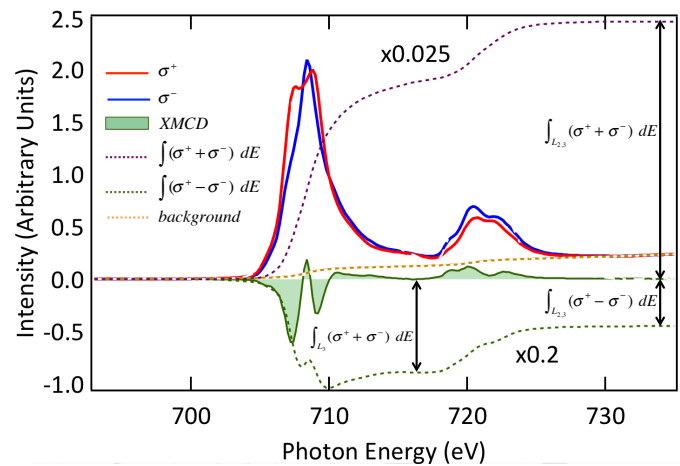

**Figure 2.** XAS and XMCD of the sample obtained under a magnetic field of 4T by normal incidence at 300K. The integrated spectra of both XMCD and total XAS saturate at ~728eV.

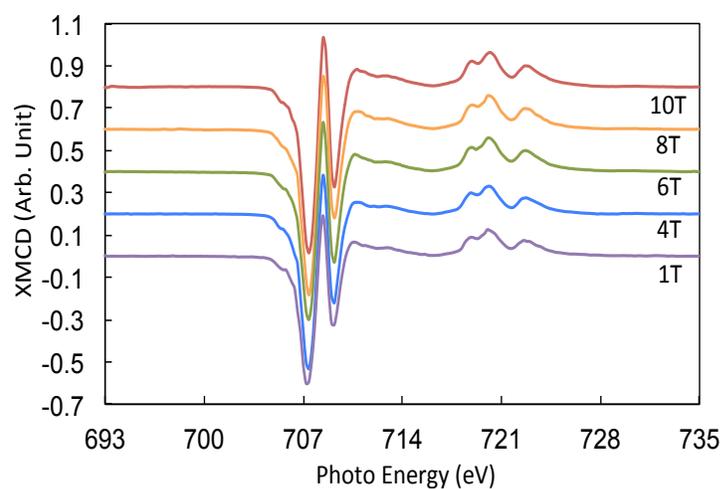

**Figure 3.** The XMCD of the sample obtained at different magnetic fields.

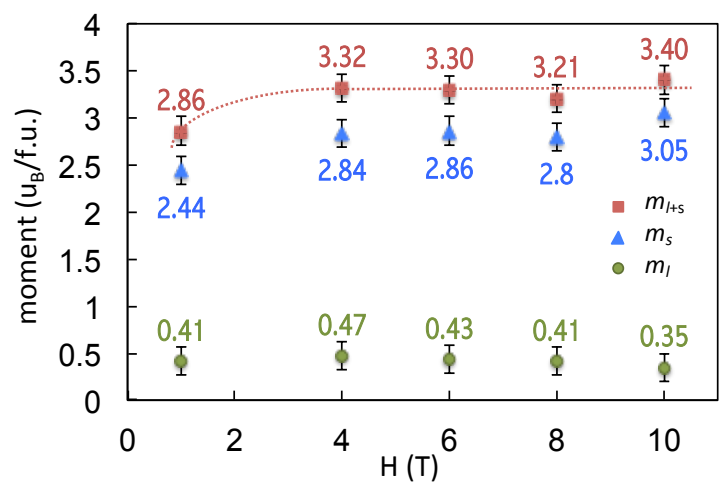

**Figure 4.** The field dependence of $m_l$, $m_s$, and $m_{l+s}$ of the sample by XMCD sum rule.